\documentstyle[epsfig,amss]{aipproc}
\newcommand{\qprime}{\mbox{$Q'~$}}
\newcommand{\qprimesq}{\mbox{$Q'^2~$}}
\newcommand{\xprime}{\mbox{$x'~$}}

\newcommand{\qprimesqx}{\mbox{$Q'^2$}}
\newcommand{\xprimex}{\mbox{$x'$}}

\newcommand{\xb}{\mbox{$x~$}}  
\newcommand{\xbx}{\mbox{$x$}}  

\newcommand{\Qsq}{\mbox{$Q^2~$}}
\newcommand{\Qsqx}{\mbox{$Q^2$}}

\newcommand{\dif}{\mbox{\rm d}}

\newcommand{\GeV}{\mbox{\rm ~GeV~}}
\newcommand{\GeVx}{\rm GeV}

\newcommand{\GeVsq}{\mbox{${\rm ~GeV}^2~$}}
\newcommand{\GeVsqx}{\mbox{${\rm ~GeV}^2$}}

\newcommand{\pb}{\mbox{${\rm ~pb}~$}}

\newcommand{\pbx}{\mbox{${\rm ~pb}$}}
\newcommand{\pbinvx}{\mbox{${\rm ~pb^{-1}}$}}

\begin{document}
\vspace{-3.0cm}
\noindent
{\tt MPI-PhE/97-13 \hfill June 1997} \\
{\tt hep-ex/9707013} \\
\vspace{-1cm}
\title{Confronting QCD Instantons with HERA Data
\footnote{talk given at the 5th Intern. Workshop on
Deep Inelastic Scattering and QCD, DIS97, Chicago 1997.} }
\author{\underline{T. Carli} and M. Kuhlen}
\address{ \vspace{-0.5cm}
Max-Planck-Institut f\"ur Physik,
Werner-Heisenberg-Institut,
F\"ohringer Ring 6, \\
D-80805 M\"unchen,
Germany,
e-mail: h01rtc@rec06.desy.de, kuhlen@desy.de 
}
\maketitle
\vspace{-0.8cm}
\begin{abstract}
The sensitivity of existing HERA data on the hadronic final
state in deep-inelastic scattering (DIS) to processes induced
by QCD instantons is systematically investigated.
The maximally allowed fraction of such processes 
in DIS is found to be on the percent level in the kinematic
domain $10^{-4} \lesssim \xb \lesssim 10^{-2}$
and $ 5 \lesssim \Qsq \lesssim 100$\GeVsqx. The best limits are 
obtained from the multiplicity distribution.
\end{abstract}
\vspace{-0.8cm}
\section{Introduction }
The
standard model contains hard processes which cannot be 
described by perturbation theory, and which violate
classical conservation laws like baryon and lepton number $B+L$
in the case of the electroweak sector and chirality in the case 
of strong interactions \cite{th:thooft}. 
Such anomalous processes are induced by instantons \cite{th:belavin}.

Observable instanton effects in electroweak processes
are only expected to be sizable in the multi-TeV region 
and are therefore out of the present experimental reach.
QCD instantons, however, could be observed  
in deep-inelastic scattering (DIS) at
HERA which collides $27.5$\GeV positrons on $820$\GeV protons
\cite{rs94,rs95,rs96,rs96b}.

The theoretical background of the cross-section calculation
and aspects of instanton phenomenology based on a Monte
Carlo simulation\cite{qcdins} for the QCD instanton
have been discussed by F. Schrempp
and A. Ringwald\cite{th:ringwalddis97,th:schremppdis97} 
at this conference.
In this contribution limits on instanton production
are derived from the most sensitive observables in the hadronic 
final state of DIS events, 
namely the
multiplicity distributions \cite{h1:mult},
the transverse energy flows \cite{h1:flow3}
and hard particle production \cite{h1:pt}.

%
%
\begin{figure}[htb]
\begin{tabular}{cc}
 \mbox{\epsfig{width=7.cm,file=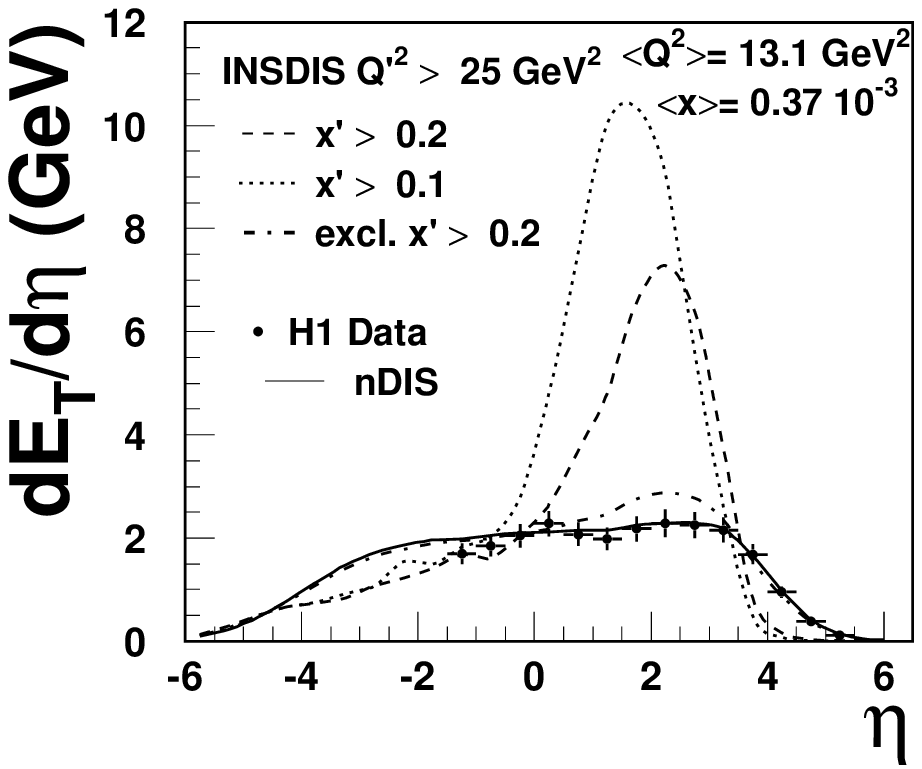}}
 \mbox{\epsfig{width=7.cm,file=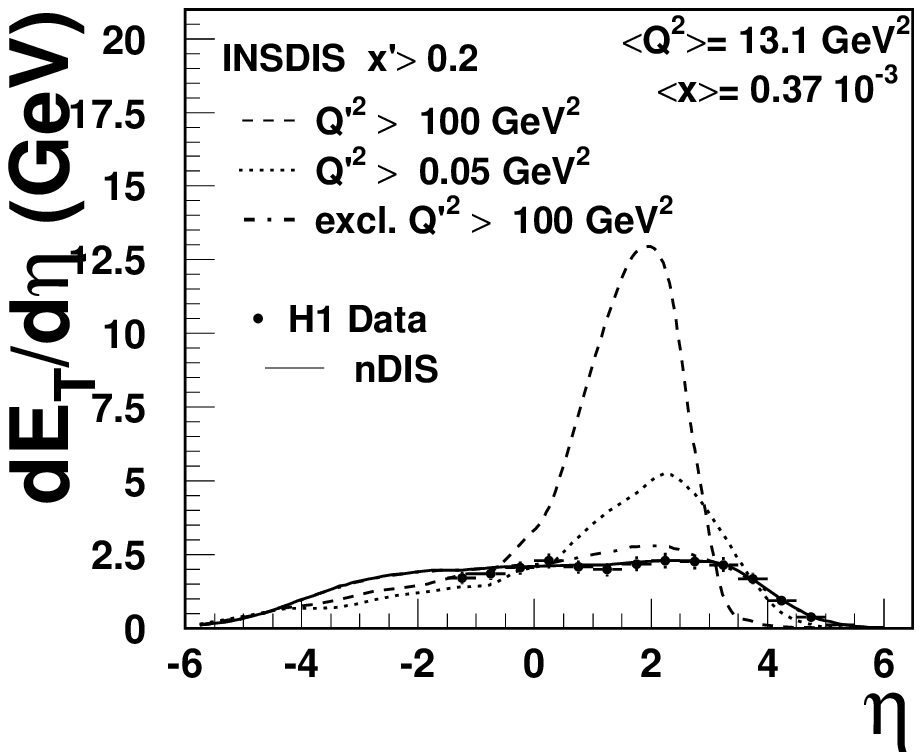}}
\end{tabular}
%
 \caption[]{ \label{fig:detdeta}
 {\it Mean transverse energy as function of the pseudo-rapidity
   in the hadronic center of mass frame.
   The proton moves to the left.
   Shown are H1 data\cite{h1:flow3}, a standard DIS prediction
   based on ARIADNE (solid line,nDIS) and the instanton expectation
   in various phase space regions in the (\xprime,\qprime) plane
   (broken lines,INSDIS). 
 }}
\end{figure}

\vspace{-0.5cm}
\section{Instanton production at HERA}

At HERA, events induced by QCD instantons predominantly 
invoke a quark-gluon fusion process. The total cross-section 
is given by a convolution of the
probability to find a gluon in the proton
$P_{g/p}$, the cross-section 
$\sigma_{q^*g}(\xprimex,\qprimesqx)$ of the instanton induced
sub-process and the probability that a photon splits
in a quark-antiquark pair in the instanton background
$P^{(I)}_{q^*/\gamma}$\cite{rs96,rs96b}.
Besides the squared transverse momentum transfer \Qsq and
the Bjorken-\xb scaling variable, this scattering process is 
characterized by \qprimesqx, the virtuality of the quark ($q^*$),
and \xprimex, the Bjorken
scaling variable associated with the $q^*g$ sub-process.
$\sigma_{q^*g}(\xprimex,\qprimesqx)$ decreases with increasing \qprimesq and
exponentially grows with decreasing \xprimex. Due to inherent
ambiguities - in particular the exponential behaviour at low \xprime - 
the calculation is only
reliable for $\xprime \gtrsim 0.2$ and $\qprimesq \gtrsim 25$\GeVsqx. 

Since the highest event rate is expected in the low-\xb regime,
DIS data down to $\xb \approx 10^{-4}$ are analyzed in this
study. When comparing to theory predictions, the
phase space is restricted to the theoretically controllable regime
by imposing limits on (\xprimex, \qprimesqx). 
The characteristics of the hadronic final state of instanton
induced events are considered to be more robust than predictions
of the absolute event rate\cite{rs96}.
To investigate the sensitivity of the event topology to 
\xprime and \qprimesqx, different instanton scenarios with
varying cut-off parameters are examined.

\vspace{-0.5cm}
\section{Experimental signature}
In its own rest frame the instanton isotropically decays into a
multi-parton state, consisting of gluons and all quark flavours
which are kinematically allowed. After fragmentation of the
semi-hard partons, a densely and uniformly populated band of particles
with characteristic flavours is expected in a certain
pseudo-rapidity ($\eta$) region. 
%
%
\begin{figure}[htb]
\begin{center}
\begin{tabular}{cc}
 \mbox{\hspace{-1.cm}
  \epsfig{width=8.cm,file=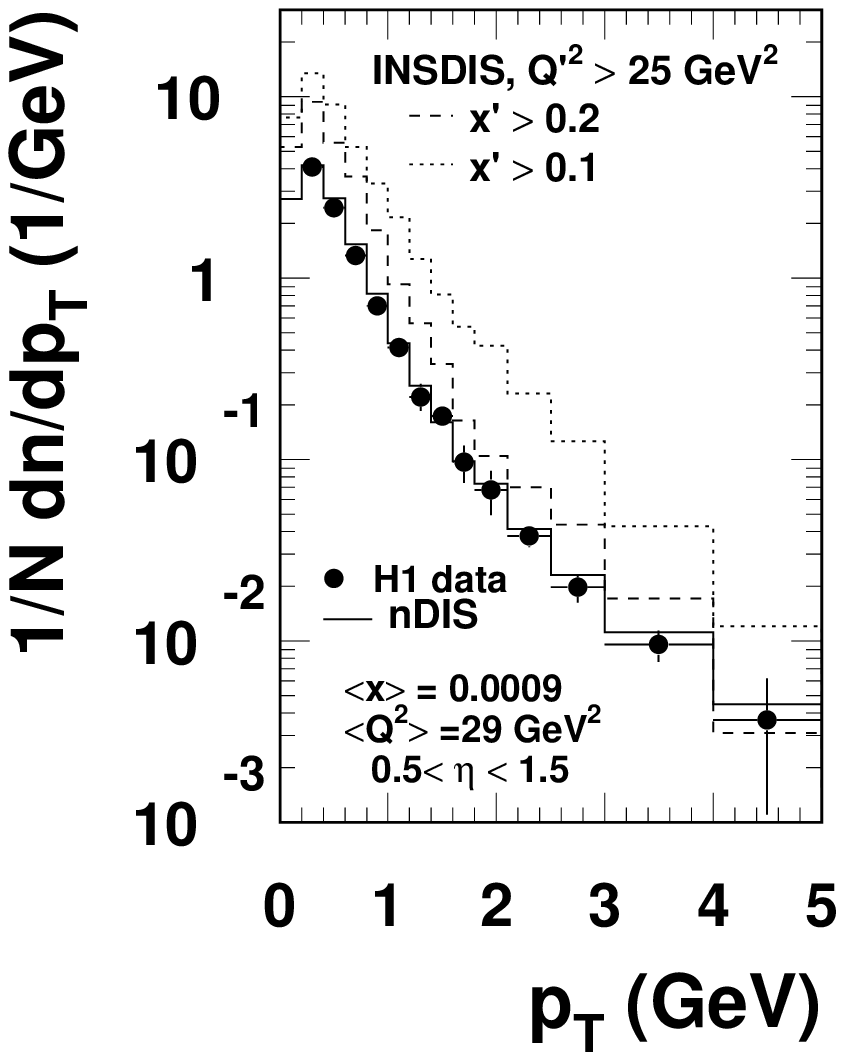,
        bbllx=28pt,bblly=424pt,bburx=321pt,bbury=747,clip=
 }}
 \mbox{\hspace{-1.cm}
  \epsfig{width=8.cm,file=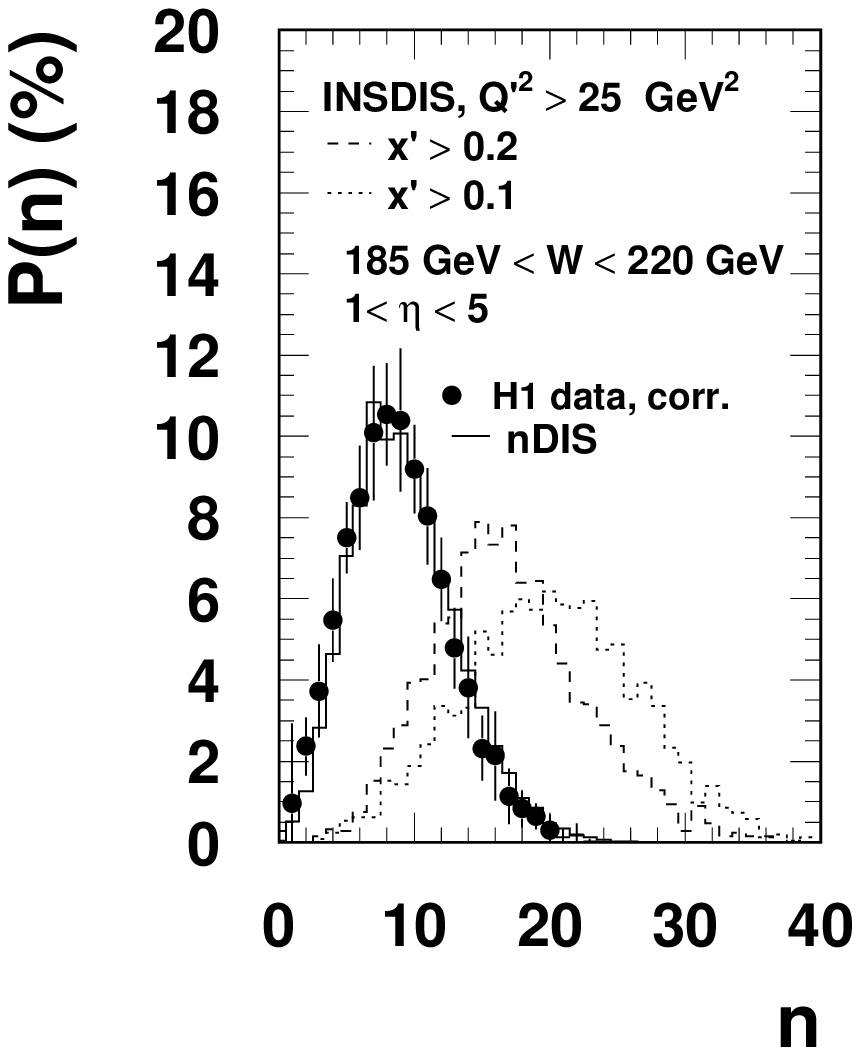,
        bbllx=28pt,bblly=424pt,bburx=321pt,bbury=747,clip=
 }}
\end{tabular}
\end{center}
%
 \vspace{-0.5cm}
 \caption[]{ \label{fig:mult}
 {\it Transverse momentum  and
  multiplicity distribution of charged particles in the hadronic
  center of mass frame.
   Shown are H1 data\cite{h1:pt,h1:mult}, a standard DIS prediction
   based on ARIADNE (solid line,nDIS) and the instanton expectation
   in different phase space region in the (\xprimex,\qprimesqx) plane
   (broken lines,INSDIS). 
 }}
\end{figure}

This band of high hadronic activity can be identified
in the mean transverse energy flow ($\dif E_T/\dif \eta$)
(see Fig.~\ref{fig:detdeta}).
While the data exhibit a plateau
of about $2$\GeV per $\eta$ unit, large mean transverse energies
of up to ${\cal O}(10~\GeVx)$ are predicted for instanton events.  
The height and the position of the instanton band depends on
the instanton mass 
$M_{\rm ins} \approx \sqrt{\qprimesqx/\xprimex}$, i.e.
for decreasing \xprime and increasing \qprimesq it gets stronger
and moves towards the proton remnant.
For large $M_{\rm ins}$ also the transverse momentum 
spectrum of charged particles becomes harder 
than for normal DIS (see Fig.~\ref{fig:mult}).

A distinct signature of instanton events is a large 
particle multiplicity. 
In Fig.~\ref{fig:mult} the distribution
of the charged particles multiplicity in $1< \eta < 5 $ 
for large $W \approx \sqrt{\Qsqx/\xbx}$ is shown. The mean of this
distribution logarithmically depends on $M^2_{\rm ins}$.

\vspace{-0.5cm}
\section{Bounds from HERA data}
%
%
\begin{figure}[htb]
\begin{center}
\begin{tabular}{cc}
 \mbox{\hspace{-0.8cm}
 \epsfig{width=10cm,file=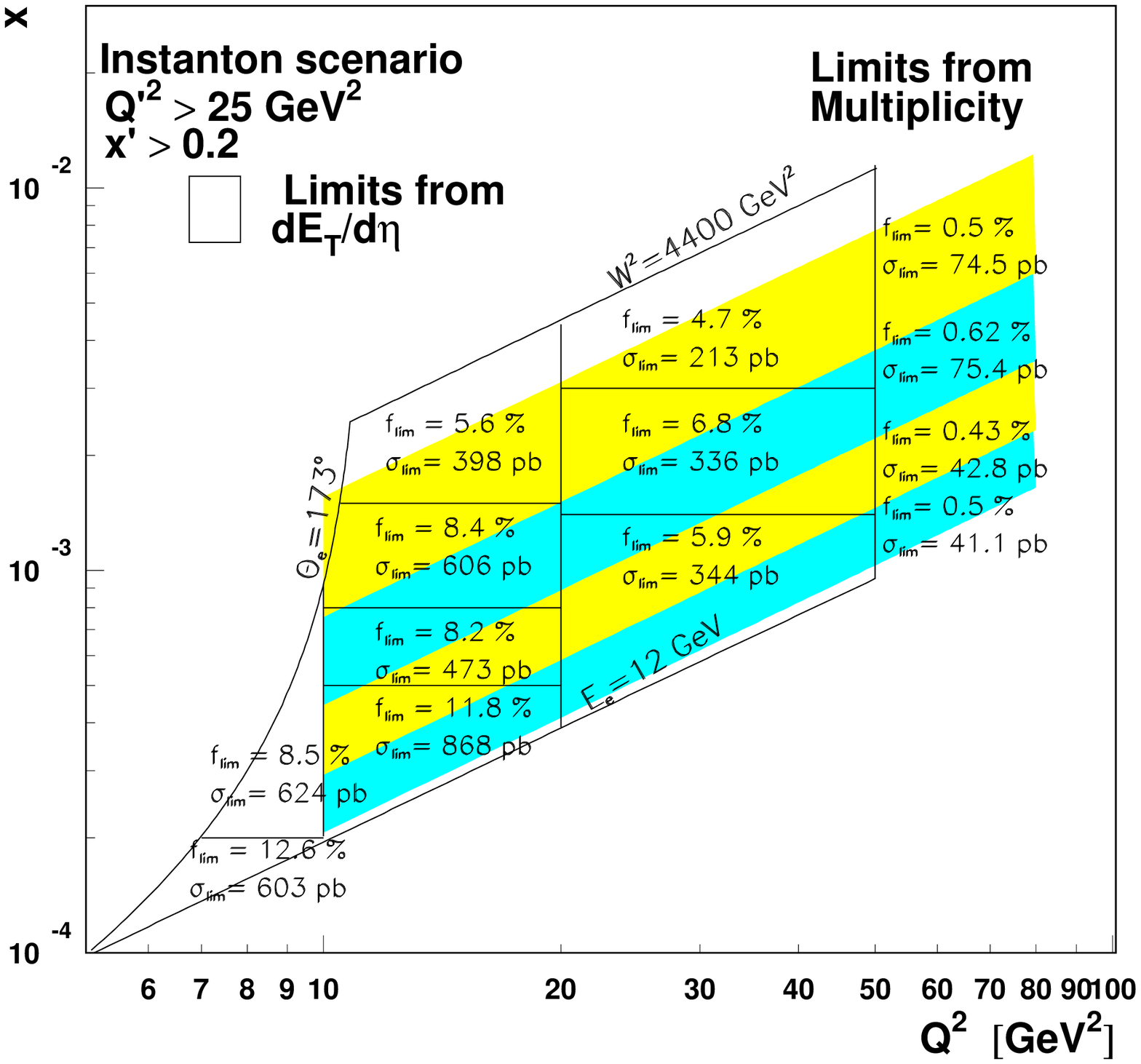,
        bbllx=28pt,bblly=142pt,bburx=561pt,bbury=668,clip=}
 }
 \mbox{\hspace{-0.45cm}
  \epsfig{width=5cm,height=10.5cm,file=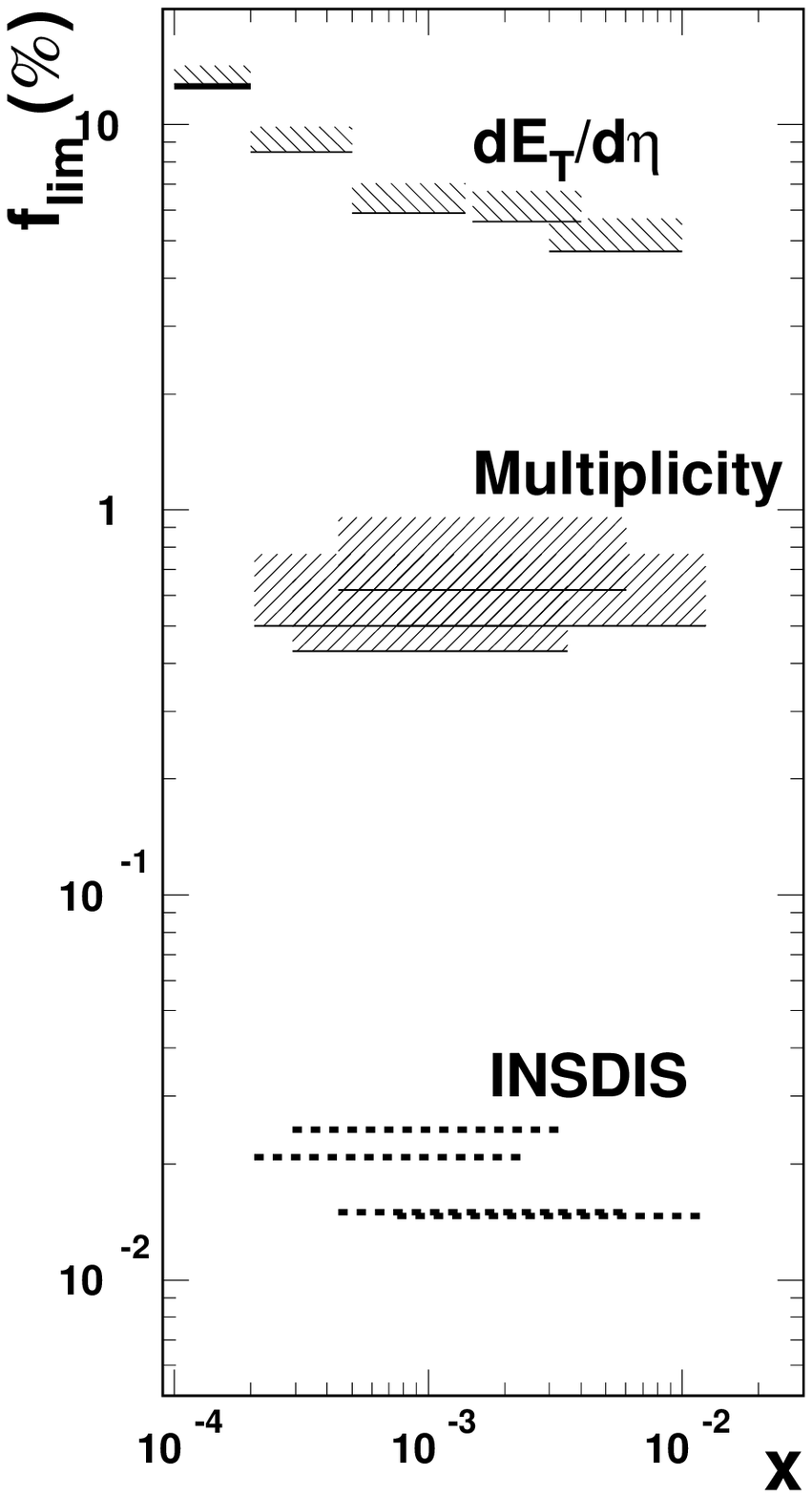}
 }
%
\end{tabular}
\end{center}
\vspace{-0.5cm}
 \caption[]{ \label{fig:limits}
 {\it Limits on instanton production with $\qprimesq > 25$\GeVsq and
      $\xprime > 0.2$ from transverse energy flows
      and the multiplicity distribution.
      a) 
      The cross-section limits ($\sigma_{\rm lim}$) 
      together with the maximally allowed instanton fraction
      $f_{\rm lim}$ in the (\xbx,\Qsqx) plane
      obtained from the $\dif E_t/\dif \eta$ (open fields)
      and multiplicity analysis (shaded fields).
      b) ($f_{\rm lim}$) versus \xb is compared to the
      theory prediction in the $W$ bin considered in 
      the multiplicity analysis (dotted line). 
 }}
\end{figure}

%
The standard DIS Monte Carlo simulation program 
ARIADNE\cite{mc:ariadne} provides
an excellent description of all available HERA data on the hadronic
final state\cite{mc:heratune}. 
Upper bounds on the cross-section
of instanton induced events ($\sigma_{\rm lim}$) in DIS can be determined
from their distinct event topology.       
Assuming that the data on $\dif E_T/ \dif \eta$ 
consist of normal DIS together with
a certain fraction of instanton events, the
maximally allowed fraction of instanton events
($f_{\rm lim}$) can be extracted using a $\chi^2$ procedure.
An example of a superposition of normal DIS with the fraction
of instanton events which can be excluded at 
$95\%$ confidence level (C.L.) is shown in Fig.~\ref{fig:detdeta}
(dashed dotted line). 
In Fig.~\ref{fig:limits} the results 
for instantons with $\xprime>0.2$ and $\qprimesq > 25$\GeVsq
are summarized (open fields).
For $ 10^{-4} \lesssim \xb \lesssim 10^{-2}$ 
and $ 10 < \Qsq < 50$\GeVsqx,  $f_{\rm lim}$ values of about
$5-10$\% can be excluded. This corresponds to cross-section limits
of $\approx 200-800$\pbx. The best limits are obtained in the domain of higher
\xb and \Qsqx.
These results can vary by at most a factor $2$
when using different choices in the details of the 
DIS model\footnote{proton structure function, hadronisation parameters etc.}.
Here, however, we used the most conservative choices.

Any dependence on the DIS model can be avoided by exploiting the
fact that no events with very high charged particle multiplicity have
been observed in the data. 
For such high multiplicies, the detection efficiency for instanton events
is still reasonably high ($8 - 20\%$), see Fig.~\ref{fig:mult}). 
At $95$\% C.L. upper limit, $f_{\rm lim}$, 
is then obtained from $f_{\rm lim} = 3/(\epsilon_i N_{DIS})$
where $\epsilon_i$ is the efficiency to detect instanton events
at large multiplicity and $N_{DIS}$ is the total number of observed DIS
events. The obtained limits in different $W$ bins are displayed
in Fig.~\ref{fig:limits} (shaded areas). For $80 < W < 220$~\GeVx,  
values $f_{\rm lim} \approx 0.5\%$ and 
$40 \lesssim \sigma_{\rm lim} \lesssim 80$\pb are obtained.

These bounds are not yet stringent enough to test the theory
which predicts an instanton fraction of 
 ${\cal O}(0.01\%)$. In the theoretically unsafe region (low \xprimex), 
where the instanton fraction is expected to be large, 
limits to constrain the theory can also be obtained.
\vspace{-0.5cm}
\section{Conclusion}
The observation of instanton effects in DIS events at HERA
would be a novel, non-perturbative manifestation of QCD
and would furthermore provide valuable indirect information about
$B+L$ violation in the multi-TeV region induced by electroweak
instantons.
The distinct event topology of instanton induced events allows
to discriminate them from normal DIS events.
Using existing HERA data on the hadronic final state 
corresponding to an integrated luminosity of ${\cal O}(1\pbinvx)$,
the maximally allowed fraction of instantons in DIS is found to 
be of ${\cal O}(1\%)$ for $ 80 < W < 220$\GeV
and $\xprime>0.2$ and $\qprimesq > 25$\GeVsqx.
The predicted instanton fraction is $\approx 0.01-0.02\%$, i.e. still below
the level excluded by existing HERA data.
The higher luminosity of ${\cal O}(10\pbinvx)$ already delivered by HERA 
and dedicated instanton searches will allow to test the absolute 
prediction of the cross-section in the near future.


%
\end{document}